\documentclass[prc,superscriptaddress,unsortedaddress,twocolumn,showpacs,preprintnumbers,amsmath,amssymb,floatfix,dvips]{revtex4}
\usepackage{amsmath}
\usepackage{amssymb}
\usepackage{times}
\usepackage{mathrsfs}
\usepackage{multirow}
\usepackage{bm}
\usepackage{ulem}
\usepackage{color}
\usepackage{graphicx}

\def\ga{\mathrel{\mathpalette\fun >}}
\def\fun#1#2{\lower3.6pt\vbox{\baselineskip0pt\lineskip.9pt
\ialign{$\mathsurround=0pt#1\hfil##\hfil$\crcr#2\crcr\sim\crcr}}}

\def\mbold#1{\mbox{\boldmath $#1$}}

\newcommand{\beq}{\begin{equation}}
\newcommand{\eeq}{\end{equation}}
\newcommand{\bea}{\begin{eqnarray}}
\newcommand{\eea}{\end{eqnarray}}

\begin{document}

\title{Simultaneous analysis of
matter radii, transition probabilities, and excitation energies
of Mg isotopes by angular-momentum-projected configuration-mixing
calculations}

\author{Mitsuhiro Shimada}
\affiliation{Department of Physics, Kyushu University, Fukuoka 819-0395, Japan}

\author{Shin Watanabe}
\affiliation{Department of Physics, Kyushu University, Fukuoka 819-0395, Japan}

\author{Shingo Tagami}
\affiliation{Department of Physics, Kyushu University, Fukuoka 819-0395, Japan}

\author{Takuma Matsumoto}
\affiliation{Department of Physics, Kyushu University, Fukuoka 819-0395, Japan}

\author{Yoshifumi R. Shimizu}
\affiliation{Department of Physics, Kyushu University, Fukuoka 819-0395, Japan}

\author{Masanobu Yahiro}
\affiliation{Department of Physics, Kyushu University, Fukuoka 819-0395, Japan}

%\date{\today}

\begin{abstract}
We perform simultaneous analysis of (1) matter radii,
(2) $B(E2; 0^+ \rightarrow 2^+ )$ transition probabilities, and
(3) excitation energies, $E(2^+)$ and $E(4^+)$,
for $^{24-40}$Mg by using the beyond mean-field (BMF) framework with
angular-momentum-projected configuration mixing
with respect to the axially symmetric $\beta_2$ deformation
with infinitesimal cranking.
The BMF calculations successfully reproduce
all of the data for $r_{\rm m}$, $B(E2)$, and $E(2^+)$ and $E(4^+)$,
indicating that it is quite useful
for data analysis, particularly for low-lying states.
We also discuss the absolute value of the deformation parameter $\beta_2$
deduced from measured values of $B(E2)$ and $r_{\rm m}$.
This framework makes it possible to investigate the effects of
$\beta_2$ deformation, the change in $\beta_2$ due to
restoration of rotational symmetry, $\beta_2$ configuration mixing,
and the inclusion of time-odd components by infinitesimal cranking.
Under the assumption of axial deformation and parity conservation,
we clarify which effect is important for each of the three measurements,
and propose the kinds of BMF calculations that are practical for each of the
three kinds of observables.
\end{abstract}

% 21.10.Gv    Nucleon distributions and halo features
% 21.60.Ev    Collective models
% 21.60.Gx    Cluster models
% 25.60.-t    Nuclear reactions, unstable-nuclei-induced
%\pacs{21.30.Fe, 24.10.Ht, 25.40.Cm, 25.55.Ci}
%21.30.Fe  Forces in hadronic systems and effective interactions
%21.65.-f  Nuclear matter
%24.10.Ht  Optical and diffraction models
%25.40.Cm  Elastic proton scattering
%25.55.Ci  Elastic and inelastic scattering (3H-,3He-, and 4He-induced reactions)

\maketitle

%Introduction
\section{Introduction}
\label{Introduction}

Recent measurements of reactions with radioactive beams have provided much
information on unstable nuclei.
Of particular interest is the island of inversion (IoI),
where the neutron number $N$ is around 20 and
the proton number $Z$ is from 10 (Ne) to 12 (Mg).
In this region, the $N=20$ magic number does not give a spherical ground state,
because a largely deformed shape is more favorable.
In fact, it has very recently been reported
as a result of measurements of total reaction cross sections
$\sigma_{\rm R}$~\cite{Takechi-Mg,Watanabe14} that
the quadrupole deformation parameter $\beta_2$ jumps up to large values
at $N=19$ for both Ne and Mg isotopes and maintains these large values up
to at least the vicinity of the neutron drip line (i.e., up to
$N=22$ for Ne isotopes and $N=26$ for Mg isotopes).
Other experiments have also shown that the low-$N$ end of the IoI is
$N=19$~\cite{Sorlin:2008jg}.
The low-$N$ end is thus rather well established,
but the high-$N$ end is still under debate and hence
the location is being intensively studied both experimentally and theoretically;
see, for example, Refs.~\cite{Hamamoto:2012mx,Door13,Koba14,Caur14}.

Rich experimental data have already been accumulated for Mg isotopes in particular. In fact, data on $\sigma_{\rm R}$ are available
for both even and odd $N$ up to $N=26$~\cite{Takechi-Mg,Watanabe14},
data for $B(E2; 0^+ \rightarrow 2^+ )$ transition probabilities
are available for even $N$ up to $N=22$~\cite{Moto95,Prit99,Chiste01,NNDC},
and data for excitation energies $E(2^+)$ and $E(4^+)$
are available for even $N$ up to $N=26$~\cite{NNDC, Door13}.
Among these three kinds of observables,
$B(E2)$ is the most useful for studying the $\beta_2$ deformation,
but it is also the most difficult to measure as demonstrated by the limited amount of data.
For this reason, $E(2^+)$ and $E(4^+)$ are often measured
instead of $B(E2)$, particularly for nuclei near the neutron
dripline. The ratio $E(4^+)/E(2^+)$ is a convenient quantity for seeing
how close nuclei are to the ideal rotor or the vibration model.

The quadrupole deformation parameter $\beta_2$ is dimensionless and
hence a convenient quantity for examining the $N$ and $Z$ dependence of nuclear
deformation, and is often estimated from
measured $B(E2; 0^+ \rightarrow 2^+ )$.
However, this estimation requires that
the root mean square (rms) matter radius
$r_{\rm m}=\sqrt{\langle r^2 \rangle}$ is properly obtained,
c.f., Eqs.~\eqref{eq:beta2} and~\eqref{eq:Q20BE2} below.
In actual data analyses, the empirical formula $1.2A^{1/3}$~[fm]
is widely used as a nuclear radius, where $A$ is the mass number.
The corresponding rms radius is
$r_{\rm m}^{\rm emp}=1.2A^{1/3}\sqrt{3/5}$~[fm],
when a uniform density is assumed.
However, it is necessary to confirm whether the empirical formula is reasonable.
In unstable nuclei it is expected that $r_{\rm m}$ is larger than
$r_{\rm m}^{\rm emp}$ because of the weakly bound nature,
and hence it is possible that the expansion effect may be misinterpreted
as an increase in $\beta_2$ deformation.
It is therefore quite important to measure the matter radius $r_{\rm m}$.

\begin{figure}[htbp]
\begin{center}
 \includegraphics[width=0.40\textwidth,clip]{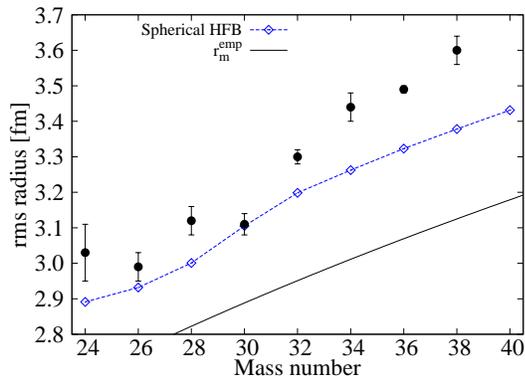}
\caption{(Color online)
$A$ dependence of matter radii for Mg isotopes.
Dots with error bars show the experimental data deduced from
the measured $\sigma_\mathrm{R}$,
where the finite-size effect of the nucleons
is subtracted from the experimental data.
The dashed line indicates the results of spherical HFB calculations.
The empirical rms radius $r_{\rm m}^{\rm emp}=1.2A^{1/3}\sqrt{3/5}$~[fm] is
also plotted as a solid line.
}
\label{rms_exp}
\end{center}
\end{figure}

The total reaction cross section $\sigma_{\rm R}$ is sensitive to
the value of $r_{\rm m}$. In fact, the values of $r_{\rm m}$ have been
deduced accurately for $^{24-38}$Mg~\cite{Watanabe14}
by using the $g$-matrix folding model~\cite{Minomo-DWS,Minomo:2011bb,Sumi:2012}
from measured $\sigma_{\rm R}$~\cite{Takechi-Mg}.
The deduced data can be regarded as experimental data for $r_{\rm m}$ because
of the accuracy of the model analyses.
The $r_{\rm m}$ values thus obtained are plotted against $A$
in Fig.~\ref{rms_exp}, where the finite-size effect of the nucleons is
subtracted from the experimental data.  The experimental values are larger than
the results (dashed line) of the spherical Hartree--Fock--Bogoliubov (HFB)
calculations with finite-range Gogny-D1S force~\cite{DeGo80,D1S}.
The difference comes from the effect of deformations, predominantly
of quadrupole type, as shown later in Sec.~\ref{Matter radii}.
This is a good example of the fact that
high-precision measurement of $\sigma_{\rm R}$ is useful for
determining the actual value of $r_{\rm m}$ including
the effect of deformations.
The empirical rms radius $r_{\rm m}^{\rm emp}=1.2A^{1/3}\sqrt{3/5}$~[fm]
is also plotted in Fig.~\ref{rms_exp} as a solid line.
The difference between the dashed and solid lines is rather large,
and we can thus conclude that the data for $B(E2)$ and $r_{\rm m}$
needs to be analyzed simultaneously in order to extract
the $\beta_2$ deformation parameter.

%%%%%%%%%%%%%%%%%%%%%%%
%%%  Figure
%%%%%%%%%%%%%%%%%%%%%%%

Measurement of $\sigma_{\rm R}$ offers the advantage
over $B(E2)$ and/or $E(2^+)$ and $E(4^+)$
that the measurement is relatively easy and possible
for all combinations of even-even, even-odd, and odd-odd nuclei.
In this paper, however, we concentrate our discussion on even-even Mg isotopes
in order to analyze all the data for $r_{\rm m}$, $B(E2)$,
and $E(2^+)$ and $E(4^+)$, simultaneously.

Taking $^{32}$Mg with $N=20$ as a representative nucleus in the IoI,
many mean-field calculations yield a spherical shape for the ground state
(see, e.g., Refs.\cite{TFH97,LFS98,RDN99}) because of the $N=20$ magicity.
As mentioned above, however,
large deformation has been reported for this nucleus by measurements of
$B(E2; 0^+ \rightarrow 2^+ )$~\cite{Moto95,Prit99,Chiste01},
$E(2^+)$ and $E(4^+)/E(2^+)$~\cite{NNDC},
and $\sigma_{\rm R}$~\cite{Takechi-Mg},
that is, $r_{\rm m}$ at $A=32$ in Fig.~\ref{rms_exp}.
In contrast, for $^{40}$Mg with $N=28$, mean-field calculations predict
a deformed ground state~\cite{TFH97,LFS98,RDN99}.
The shell closure is thus more fragile for $N=28$ than
for $N=20$~\cite{Bas07,Take12}.

Very recently, state-of-the-art calculations~\cite{RER15} have been performed
for $^{24-34}$Mg with $N=12$ -- $22$ in the beyond mean-field framework (BMF)
with angular-momentum-projected configuration mixing
for the triaxial quadrupole deformation and the cranking frequency
by employing the finite-range Gogny D1S force.
In this framework,
a set of intrinsic mean-field states is first obtained by minimizing
the particle-number projected HFB energy with fixed deformation parameters
$(\beta_2,\gamma)$ and with the rotational frequency $\omega_{\rm rot}(J_c)$
that gives the average angular momentum $\langle J_x \rangle=\sqrt{J_c(J_c+1)}$.
The HFB states specified by $(\beta_2,\gamma,J_c)$ are
then configuration-mixed after the angular momentum projection (AMP)
within the generator coordinate method.
The state-of-the-art calculations predict a deformed ground state for $^{32}$Mg,
and reproduce measured excitation energies well.
Similar systematic calculations have also been performed within
the relativistic framework in Ref.~\cite{YMC11},
although the cranking prescription was not taken into account.

In this paper, we analyze all of the observables $r_{\rm m}$, $B(E2)$,
and $E(2^+)$ and $E(4^+)$ simultaneously for $^{24-40}$Mg with $N=12$ -- $28$
by comparing theoretical results with experimental data.
We use the BMF framework of the angular-momentum-projected
configuration mixing with respect to the deformation.
We also apply infinitesimal cranking~\cite{TS16}, which has recently been shown
to give a practical and good description of low-lying rotational states.
In the present calculations,
we consider only the axially symmetric deformation for the configuration mixing,
since antisymmetrized molecular dynamics (AMD) shows that
the average triaxiality $\gamma$ is zero or small
for $^{24-40}$Mg~\cite{Watanabe14}
and the state-of-the-art BMF calculation in
Ref.~\cite{RER15} has also shown that the energy quickly increases
as the triaxial deformation increases.
In fact, the present BMF calculations agree well with the experimental data
and with the results of Ref.~\cite{RER15}
for $E(2^+)$ and $E(4^+)$ in $^{24-34}$Mg,
as shown later in Sec.~\ref{Excitation energies}.
The present BMF framework is thus considered rather reliable.
It is therefore of interest to apply
the present BMF framework to the simultaneous analysis
of the three kinds of observables to confirm
the reliability of the theoretical framework.
The main result of this paper is that
the present BMF framework successfully reproduces all of the data
for $r_{\rm m}$, $B(E2)$, and $E(2^+)$ and $E(4^+)$.

The present BMF framework allows us to investigate the
following four effects separately:
\begin{enumerate}
\renewcommand{\labelenumi}{(\roman{enumi})}
\item The $\beta_2$ deformation
\item The change in $\beta_2$ due to the restoration of rotational symmetry
by the AMP, that is, the change in $\beta_2$
from a minimum on the HFB energy surface to a minimum
on the projected-energy surface
\item The configuration mixing for the axially symmetric $\beta_2$ deformation
\item The infinitesimal cranking for AMP,
that is, the inclusion of time-odd components
in the mean-field wave function by the cranking procedure
\end{enumerate}
The effects (ii) to (iv) are correlations we take beyond the HFB calculations.
We clarify the effect that is important for each of the three observables
and propose the kinds of BMF calculations that are practical for each of the
three kinds of observables.

We explain the present theoretical framework in Sec.~\ref{Theoretical framework}.
The results of our BMF calculations are shown compared
with experimental data in Sec.~\ref{Results}.
The absolute values of the deformation parameter $|\beta_2|$
extracted from the measured $B(E2)$ and $r_{\rm m}$ are also discussed
in more detail in the Appendix.
Section~\ref{Summary} summarizes this work.

%Theoretical framework
\section{Theoretical framework}
\label{Theoretical framework}

The wave function of the angular-momentum-projected
configuration-mixing approach~\cite{RS80} is introduced generally as
\begin{equation}
 | \Psi^I_{M} \rangle =
 \sum_{K n} g^I_{K n} \hat P^I_{MK}|\Phi_n \rangle ,
\label{eq:prjc}
\end{equation}
where $\hat P^I_{MK}$ is the angular momentum projector and
$|\Phi_n \rangle$ ($n=1,2,\cdots,N$) is the set of mean-field states
given below.
The coefficients $g^I_{K n}$ are determined
by solving the following Hill-Wheeler equation
\begin{equation}
 \sum_{K^\prime n^\prime }{\cal H}^I_{Kn,K^\prime n^\prime }\ g^I_{K^\prime n^\prime } =
 E_I\,
 \sum_{K^\prime n^\prime }{\cal N}^I_{Kn,K^\prime n^\prime }\ g^I_{K^\prime n^\prime },
\end{equation}
where the Hamiltonian and norm kernels are defined by
\begin{equation}
 \left\{ \begin{array}{c}
   {\cal H}^I_{Kn,K^\prime n^\prime } \\ {\cal N}^I_{Kn,K^\prime n^\prime } \end{array}
 \right\} = \langle \Phi_n |
 \left\{ \begin{array}{c}
   \hat{H} \\ 1 \end{array}
 \right\} \hat{P}_{KK^\prime }^I | \Phi_{n'} \rangle.
\end{equation}

The mean-field states used in Eq.~(\ref{eq:prjc}) are prepared
as a function of the dimensionless parameter $\beta_2$ for
the axially symmetric quadrupole deformation defined by
\begin{equation}
 \beta_2 = \frac{4\pi}{5} \frac{Q_{20}}{A\langle r^2 \rangle},
\label{eq:beta2}
\end{equation}
where the mass quadrupole moment $Q_{20}$
and the mean square radius $\langle r^2 \rangle$ are
calculated by the mean-field wave function $|\Phi(\beta_2) \rangle$
\begin{eqnarray}
 Q_{20} &=& \langle \Phi | \sum_{a=1}^A(r^2 Y_{20})_a | \Phi \rangle,
\label{eq:Qmom} \\
 A\langle r^2 \rangle &=& \langle \Phi | \sum_{a=1}^A(r^2)_a | \Phi \rangle.
\label{eq:rms}
\end{eqnarray}
In the actual calculation, the set of the $\beta_2$ values is
properly chosen and the configuration mixing is performed for
$|\Phi_n \rangle=|\Phi(\beta_2^{(n)})\rangle$ $(n=1,2,\cdots,N_{\beta_2})$,
which are obtained by the constrained HFB calculation
using the quadrupole operator $r^2Y_{20}$ as a constraint.
The augmented Lagrangian method~\cite{Stas10} is employed to achieve
the desired value of the constraint.

In axially symmetric deformation, only the $K=0$ components
survive in Eq.~(\ref{eq:prjc}).
However, it is known that the moment of inertia for rotational excitation
is underestimated if this kind of time-reversal invariant mean-field state
is utilized for the AMP calculation.
The time-odd components in the HFB wave function
are important for increasing the moment of inertia~\cite{TS12}.
An efficient way to include the time-odd components is the cranking method; that is, the cranked HFB state $|\Phi_{\rm cr}(\beta_{2}^{(n)}) \rangle$
is calculated using the cranked Hamiltonian
\begin{equation}
\hat{H}'=\hat{H}-\omega_{\rm rot}J_y,
\label{eq:Hcr}
\end{equation}
by replacing the original $\hat{H}$.
The axis of rotation is chosen to be the $y$-axis,
which is perpendicular to the symmetry axis ($z$-axis).
The so-called cranking term, $-\omega_{\rm rot}J_y$, breaks
the time-reversal symmetry of the wave function and includes
the Coriolis and centrifugal force effects.
That is, the $K$-mixing induced by the cranking term
affects the excited $I>0$ states in the AMP calculations.
It has been shown that the small cranking frequency
$\omega_{\rm rot}$ is enough to increase the moment of inertia
and the result is independent of the actual value of $\omega_{\rm rot}$
used as long as it is small.
This method is called infinitesimal cranking~\cite{TS16}.
Note that the mean-field and $0^+$ AMP calculations
are not affected by infinitesimal cranking.

We have also studied configuration mixing
with respect to the cranking frequency,
and have shown that the spin-dependence of high-spin moments of inertia
can be well described by superposing the angular-momentum-projected HFB
states with various cranking frequencies~\cite{STS15,STS16}.
However, infinitesimal cranking is sufficient for low-spin states
like the first excited $2^+$ and $4^+$ states considered here,
and we therefore employ it instead of configuration mixing
for the cranking frequency as in Ref.~\cite{RER15}.
Although we use $\hbar\omega_{\rm rot}=10$ keV for the actual value of
infinitesimal cranking in the following calculations,
the result does not depend on it.

Since the functions $|\Phi_n \rangle$ are not orthogonal,
$g^I_{Kn}$ cannot be treated as probability amplitudes.
We thus introduce the properly normalized amplitude~\cite{RS80}
\begin{equation}
 f^I_{K n} = \sum_{K^\prime n^\prime} (\sqrt{{\cal N}})^I_{Kn,K^\prime n^\prime } g^I_{K^\prime n^\prime}.
\end{equation}
That is, the probability of $n$-th HFB states $|\Phi(\beta_{2}^{(n)}) \rangle$
in Eq.~(\ref{eq:prjc}) is given by
\begin{equation}
 p^I(\beta_{2}^{(n)})=\sum_{K}|f^I_{K n}|^2.
 \label{eq:probq20}
\end{equation}
The rms deformation parameter, for example,
is calculated by using this probability
\begin{equation}
\bar{\beta}_2=\langle (\beta_2)^2 \rangle^{1/2},
 \label{eq:beta2m}
\end{equation}
with
\begin{equation}
 \langle (\beta_2)^2 \rangle=
 \sum_n\bigl(\beta_{2}^{(n)}\bigr)^2\,p^I(\beta_{2}^{(n)}).
 \label{eq:beta2av}
\end{equation}

Once the set of cranked HFB states,
$|\Phi_{\rm cr}(\beta_{2}^{(n)}) \rangle$; $n=1,2,\cdots,N_{\beta_2}$,
is thus obtained and the amplitudes $g^I_{Kn}$ in Eq.~(\ref{eq:prjc})
are determined, it is straightforward~\cite{RS80} to calculate the rms radius,
$[\langle\Psi^I_M|\sum_{a=1}^A (r^2)_a|\Psi^I_M\rangle]^{1/2}$,
and the $E2$ transition probability, $B(E2)$,
in addition to the energy eigenvalue $E_I$.
In actual calculations, we use the harmonic-oscillator basis expansion with
the frequency $\hbar\omega=41/A^{1/3}$~MeV, and retain all the basis states
with the oscillator quantum numbers $(n_x,n_y,n_z)$ satisfying
$n_x+n_y+n_z \le N_{\rm osc}^{\rm max}=8$.
In other words, we include the nine major shells.
The numbers of mesh points for the numerical integration with respect to
the Euler angles ($\alpha,\beta,\gamma$) in the angular momentum projector
are taken to be $N_\beta=42$ and $N_\alpha=N_\gamma=10$,
which are sufficient for the low-spin states of
essentially axially symmetric nuclei with infinitesimal cranking.
We adopt the Gogny-D1S parameter set~\cite{D1S} for the effective interaction.

%%%%%%%%%%%%%%%%%%%%%
\begin{figure}[!htb]
\begin{center}
 \includegraphics[width=0.40\textwidth]{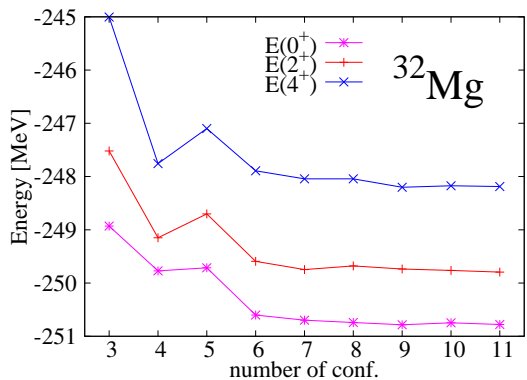}
\end{center}
\vspace{-12pt}
\caption
{(Color online)
Convergence of the $\beta_2$ configuration-mixing calculations
for the $0^+$, $2^+$, and $4^+$ energies as the number
of HFB states with different values of quadrupole moment
chosen equidistantly in the range,
$-80$ [fm$^2]$ $\le Q_{20} \le$ 140 [fm$^2]$, is increased.
}
\label{fig:E024conv}
\end{figure}
%%%%%%%%%%%%%%%%%%%%

In the actual calculations of $\beta_2$ configuration mixing,
the quadrupole moment in Eq.~(\ref{eq:Qmom}) is constrained.
Constraining the $\beta_2$ value is nontrivial because it depends
on both the quadrupole moment and the radius.
Figure~\ref{fig:E024conv} shows an example of convergence
for the configuration-mixing calculation for $^{32}$Mg
as a function of the number of configurations.
It can be seen that the results are stable for $N_{\beta_2} \ga 8$,
and thus we take $N_{\beta_2}=10$ in the following calculations.
The deformation parameters of the HFB states employed
are in the range of $ -0.5 \lesssim \beta_2 \lesssim 0.8$,
which mostly covers the important low-energy region of
the potential energy curve, c.f. Sec.~\ref{Energy surfaces}.
The sets of values, ($\beta_2^{(n)};n=1:10$), for the calculated Mg isotopes
are shown in Table~\ref{table:betaconf}.

%%%%%%%%%%%%%%%%%%%%%%%%%%%%%%
\renewcommand{\arraystretch}{1.25}
\begin{table}[!htb]
\begin{center}
\begin{tabular}{c c c c c c c c c c c}
\hline \hline
 Nuclide  &  \multicolumn{10}{c}{$\beta_2^{(n)}; n=1:10$}  \\
\hline
 $^{24}$Mg  & $-$0.45   & $-$0.32 & $-$0.15 & 0.04 & 0.23 & 0.41& 0.55 & 0.66 &0.75 &0.83 \\
 $^{26}$Mg  & $-$0.50   & $-$0.40 & $-$0.26 & 0.07 & 0.12 & 0.30& 0.45 & 0.58 &0.69 &0.78 \\
 $^{28}$Mg  & $-$0.49   & $-$0.37 & $-$0.21 & 0.02 & 0.19 & 0.37& 0.52 & 0.65 &0.75 &0.83 \\
 $^{30}$Mg  & $-$0.52   & $-$0.41 & $-$0.27 & 0.09 & 0.11 & 0.30& 0.47 & 0.60 &0.71 &0.80 \\
 $^{32}$Mg  & $-$0.49   & $-$0.37 & $-$0.23 & 0.05 & 0.13 & 0.31& 0.46 & 0.60 &0.70 &0.79 \\
 $^{34}$Mg  & $-$0.49   & $-$0.38 & $-$0.23 & 0.05 & 0.14 & 0.32& 0.48 & 0.60 &0.72 &0.82 \\
 $^{36}$Mg  & $-$0.50   & $-$0.38 & $-$0.23 & 0.04 & 0.15 & 0.33& 0.48 & 0.61 &0.73 &0.82 \\
 $^{38}$Mg  & $-$0.50   & $-$0.39 & $-$0.24 & 0.06 & 0.13 & 0.31& 0.46 & 0.59 &0.70 &0.79 \\
 $^{40}$Mg  & $-$0.50   & $-$0.38 & $-$0.24 & 0.05 & 0.14 & 0.32& 0.46 & 0.60 &0.70 &0.79 \\
\hline \hline
\end{tabular}
\caption
{
Parameter sets ($\beta_2^{(n)}; n=1:10$) taken for $\beta_2$
configuration mixing.
}
\label{table:betaconf}
\end{center}
\end{table}
%%%%%%%%%%%%%%%%%%%%%%%%%%%%%%

The Hill--Wheeler equation suffers from the numerical problem of
vanishing norm states~\cite{RS80}.
To avoid this, the eigenstates of the norm
kernel whose eigenvalues are smaller than a certain value need to be excluded.
This norm cutoff value needs to be as small as possible, and
we normally set it to $10^{-10}$.  However, we have found that this value is
too small for calculations including $\beta_2$ configuration mixing with
infinitesimal cranking in the nuclei $^{24,26}$Mg, for which
we use the larger value of $10^{-7}$.

\section{Results}
\label{Results}

We performed the following four kinds of
beyond mean-field (BMF) calculations
in addition to spherical and deformed HFB calculations
in order to separately investigate the following effects:
(i) $\beta_2$ deformation,
(ii) restoration of the rotational symmetry by
the AMP,
(iii) $\beta_2$ configuration mixing (CM$\beta_2$), and
(iv) infinitesimal cranking (CR).

\begin{enumerate}
\item BMF(AMP+CM$\beta_2$+CR):
This is the full calculation that includes the effects of (i) to (iv)
with the wave function
\begin{equation}
 | \Psi^I_{M} \rangle =
 \sum_{K n} g^I_{K n} \hat P^I_{MK}|\Phi_{\rm cr}(\beta_2^{(n)}) \rangle .
\label{eq:wf1}
\end{equation}
\item BMF(AMP+CM$\beta_2$):
This BMF calculation includes the effects of (i) to (iii)
with the wave function
\begin{equation}
 | \Psi^I_{M} \rangle =
 \sum_{n} g^I_{0 n} \hat P^I_{M0}|\Phi(\beta_2^{(n)}) \rangle .
\label{eq:wf2}
\end{equation}
\item BMF(AMP+CR):
This BMF calculation includes the effects of (i), (ii) and (iv)
with the wave function
\begin{equation}
 | \Psi^I_{M} \rangle =
 \sum_{K} g^I_{K} \hat P^I_{MK}|\Phi_{\rm cr}(\beta_2^{\rm min}) \rangle ,
\label{eq:wf3}
\end{equation}
where $\beta_2^{\rm min}$ is the value of $\beta_2$ which gives
the minimum energy of the projected $0^+$ ground state.
\item BMF(AMP):
This BMF calculation includes the effects of (i) and (ii)
with the wave function
\begin{equation}
 | \Psi^I_{M} \rangle =
  g^I_{0} \hat P^I_{M0}|\Phi(\beta_2^{\rm min}) \rangle ,
\label{eq:wf4}
\end{equation}
where the $\beta_2^{\rm min}$ value is the same as
in Eq.~(\ref{eq:wf3}) because the cranking frequency is infinitesimally small.
\end{enumerate}

BMF(AMP+CM$\beta_2$) calculations were first performed for the axially symmetric deformation
using the Gogny D1S force in Ref.~\cite{RER02}.
Although we confirmed their results,
we present our BMF(AMP) and BMF(AMP+CM$\beta_2$)
results in addition to BMF(AMP+CM$\beta_2$+CR) and BMF(AMP+CR) to aid the understanding in our discussion.

Number projection is not performed in the present work,
and number conservation is treated approximately~\cite{BDF89}
by replacing the Hamiltonian
$H \rightarrow H-\lambda_\nu (N-N_0)-\lambda_\pi (Z-Z_0)$,
where $N_0$ and $Z_0$ are the neutron and proton numbers to be fixed,
and the neutron and proton chemical potentials
$\lambda_\nu$ and $\lambda_\pi$ are chosen to be
those of the first ($n=1$) HFB state.
We checked the expectation values of the numbers
in the full BMF(AMP+CM$\beta_2$+CR) calculations.
The average deviations $|\langle  N-N_0  \rangle|$
and $|\langle  Z-Z_0  \rangle|$ for the calculated cases are typically
$0.02-0.08$, and the worst case is 0.19 for neutrons in $^{38}$Mg,
which is still less than 1\% of $N_0$.  Therefore the number conservation
on average is maintained well in the configuration-mixing calculations.
The neutron or proton pairing correlation vanishes depending on
the quadrupole moment, or $\beta_2$ (see, e.g., Fig.~3 of Ref.~\cite{RER02}).
The number fluctuations
$\langle\Delta N^2\rangle$ and $\langle\Delta Z^2\rangle$ for the HFB states
with non-vanishing pairing correlations are typically about $1.4-4$ for neutrons
and about $1.2-3$ for protons.
Thus, the pairing correlations are not very strong for these Mg isotopes,
and variation after number projection may be necessary for better
treatment of the pairing correlation,
which is outside the scope of this work.

\subsection{Energy surfaces}
\label{Energy surfaces}

\begin{figure}[!htb]
\begin{center}
 \includegraphics[width=0.40\textwidth]{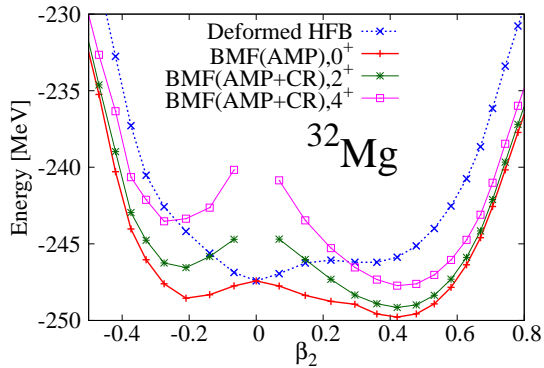}
\end{center}
\vspace{-12pt}
\caption
{(Color online)
Potential-energy curves by deformed HFB and BMF(AMP+CR) calculations
for $^{32}$Mg with the Gogny-D1S interaction.
The ground-state ($I^\pi = 0^+$) energy as well as
excited $2^+$ and $4^+$ energies are plotted as a function of $\beta_2$.
}
\label{fig:Esrf-beta32Mg}
\end{figure}

Figure~\ref{fig:Esrf-beta32Mg} shows the ground-state potential energy curves
obtained by deformed HFB (dashed line) and by BMF(AMP) (solid line) calculations
for $^{32}$Mg.
The excited $2^+$ and $4^+$ states from the BMF(AMP+CR) calculations
are also included (the deformed HFB and the BMF(AMP) $0^+$ state
are not affected by infinitesimal cranking).
Compared to the similar calculation in Ref.~\cite{RER02}
which did not include infinitesimal cranking, the $2^+$ and $4^+$ energy curves
are considerably lower in energy, which shows the importance of
the effect of time-odd components in the wave function.
Although deformed-HFB calculations yield a minimum at $\beta_2=0$,
large energy gains are obtained by BMF(AMP) calculations for finite $\beta_2$.
Consequently, BMF(AMP) calculations suggest that a considerably
large prolate deformation ($\beta_2 \approx $ 0.42) is favored,
which is consistent with the suggestion
by the experiment in Refs.~\cite{Moto95,Prit99,Chiste01,NNDC}.
Thus, it is quite important to perform AMP
to obtain the correct value of nuclear deformation.
The spherical barrier, which corresponds to the energy difference between
the prolate and spherical states, is not very large ($\Delta E \approx 2.3$ MeV).
An oblate minimum is found at $\beta_2 \approx -0.23$,
and the energy difference between the prolate and oblate states
is $\Delta E \approx 870$ keV.
These results indicate that this nucleus is soft
with respect to $\beta_2$ deformation and that configuration mixing should therefore be taken into account.

%%%%%%%%%%%%%%%%%%%%%
%%%%%%%%%%%%%%%%%%%%

\begin{figure}[!htb]
\begin{center}
 \includegraphics[width=0.40\textwidth]{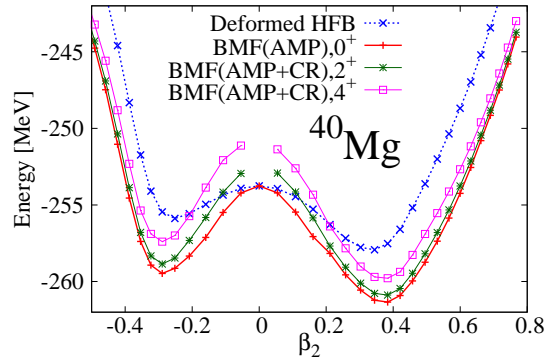}
\end{center}
\vspace{-12pt}
\caption
{
(Color online)
Same as Fig.~\ref{fig:Esrf-beta32Mg}, but for $^{40}$Mg.
}
\label{fig:Esrf-beta40Mg}
\end{figure}

Figure~\ref{fig:Esrf-beta40Mg} shows the potential energy curves for $^{40}$Mg.
Again, the $2^+$ and $4^+$ energies are considerably lower in energy
compared to Ref.~\cite{RER02}.
Prolate deformation is favored in both deformed HFB and BMF(AMP) calculations.
In contrast to the case of $^{32}$Mg, a considerably deep prolate minimum is
obtained.
The spherical barrier is large ($\Delta E \approx 5.7$ MeV).
An oblate minimum is found at $\beta_2 \approx -0.38$, and the energy difference between the prolate and oblate states is $\Delta E \approx 1.8$ MeV.
Thus, the effects of $\beta_2$ configuration mixing
may be small for $^{40}$Mg.

%%%%%%%%%%%%%%%%%%%%%
%%%%%%%%%%%%%%%%%%%%

\begin{figure}[!htb]
\begin{center}
 \includegraphics[width=0.45\textwidth]{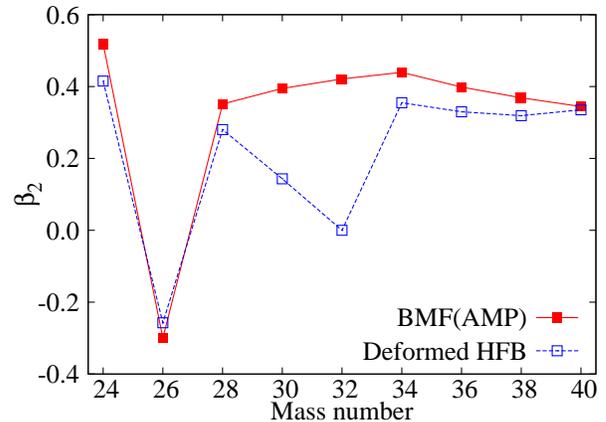}
\end{center}
\caption
{(Color online)
Deformation parameters $\beta_2$ that yield
energy minima in deformed-HFB and BMF(AMP) calculations versus mass number.
}
\label{fig:betahfbprjc}
\end{figure}

We determine nuclear deformation from the $\beta_2$ value
that yields an energy minimum.
Figure~\ref{fig:betahfbprjc} shows this value plotted against
$A$ for deformed-HFB calculations (open squares) and BMF(AMP)
calculations for the $0^+$ ground states (closed squares).
The two results are quite different for $^{30, 32}$Mg.
In the BMF(AMP) calculations,
these $\beta_2^{\rm min}$ values are used for the HFB states,
c.f. Eqs.~(\ref{eq:wf3}) and~(\ref{eq:wf4}).
This shows that it is essential to determine it properly by calculations that restore
the rotational symmetry,
particularly around $A=32$ ($N=20$).

%%%%%%%%%%%%%%%%%%%%%
%%%%%%%%%%%%%%%%%%%%

\subsection{Matter radii}
\label{Matter radii}

\begin{figure}[!htb]
\begin{center}
 \includegraphics[width=0.45\textwidth]{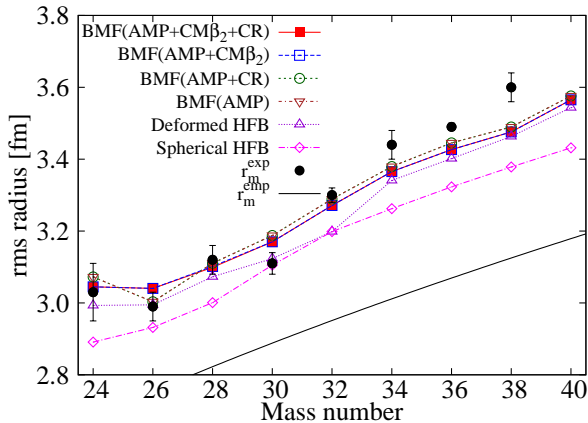}
\end{center}
\vspace{-12pt}
\caption
{(Color online)
Root mean square radii of Mg isotopes.
Four results from BMF(AMP+CM$\beta_2$+CR), BMF(AMP+CM$\beta_2$), BMF(AMP+CR),
and BMF(AMP) are shown by closed squares, open squares, open circles,
open down-triangles, respectively.
For comparison, deformed and spherical HFB results are also shown
by open triangles and open diamonds, respectively.
Experimental radii $r_{\rm m}^{\rm exp}$ are deduced from measured
$\sigma_{\rm R}$~\cite{Watanabe14,Takechi-Mg}.
The solid line shows the empirical formula
$r_{\rm m}^{\rm emp}=1.2A^{1/3}\sqrt{3/5}$~[fm] for the rms matter radius
in the spherical limit that is obtained from
a uniform density with a radius $1.2A^{1/3}$~[fm].
}
\label{fig:rmsr}
\end{figure}

Figure~\ref{fig:rmsr} shows the rms matter radius $r_{\rm m}$
for the ground-state densities of Mg isotopes from $A=24$ to 40
as obtained by analysis of
the reaction cross sections~\cite{Takechi-Mg,Watanabe14}.
The finite-size effect of nucleons is subtracted from the experimental data.
In the calculation, the center-of-mass correction has been performed
by replacing the nucleon coordinate $\mbold{r}$ with $\mbold{r}-\mbold{R}$
where \mbold{R} is the center-of-mass coordinate.
Although this effect is small at about 1\%, it is not completely negligible.
The difference between deformed-HFB results (open triangles)
and spherical-HFB results (open diamonds) is quite large,
indicating that nuclear deformation plays an important role in $r_{\rm m}$.
Our full BMF(AMP+CM$\beta_2$+CR) calculations
(closed squares) yielded excellent
agreement with the experimental data $r_{\rm m}^{\rm exp}$,
compared to the deformed-HFB results.
The four BMF results, that is, BMF(AMP+CM$\beta_2$+CR),
BMF(AMP+CM$\beta_2$), BMF(AMP+CR), and BMF(AMP), are all similar.
The non-negligible enhancement
from deformed-HFB results to BMF(AMP+CM$\beta_2$+CR) results,
particularly for $^{30,32}$Mg, is thus thought to come mainly from
the large change in the equilibrium deformation caused by the AMP
as shown in Fig.~\ref{fig:betahfbprjc}.
It should be noted that all of the multipole deformations
contribute to increasing the nuclear radius.
Although the effect of the quadrupole deformation is dominant
in the present study of Mg isotopes,
the hexadecapole deformation is non-negligible for most of the isotopes.

%%%%%%%%%%%%%%%%%%%%%
%%%%%%%%%%%%%%%%%%%%

The empirical formula $1.2 A^{1/3}$~[fm] is widely used
for nuclear radii.   The corresponding rms radius
is also plotted as a solid line, which shows that
the formula largely underestimates $r_{\rm m}^{\rm exp}$.
The matter densities are thus much more
extended than typical nuclear densities with radii $1.2 A^{1/3}$~[fm]
for even stable Mg isotopes.
If the deformation parameter $\beta_2$ is deduced from measured quantities,
it is necessary that $r_{\rm m}^{\rm exp}$ is available,
as shown in Eq.~(\ref{eq:beta2}).
At least for lighter nuclei such as Mg isotopes,
the empirical formula $r_{\rm m}^{\rm emp}=1.2 A^{1/3} \sqrt{3/5}$~[fm]
should not be used instead of $r_{\rm m}^{\rm exp}$,
even if it is not available.
This is discussed in Sec.~\ref{Transition probabilities}.

\subsection{Transition probabilities}
\label{Transition probabilities}

\begin{figure}[!htb]
\begin{center}
 \includegraphics[width=0.45\textwidth]{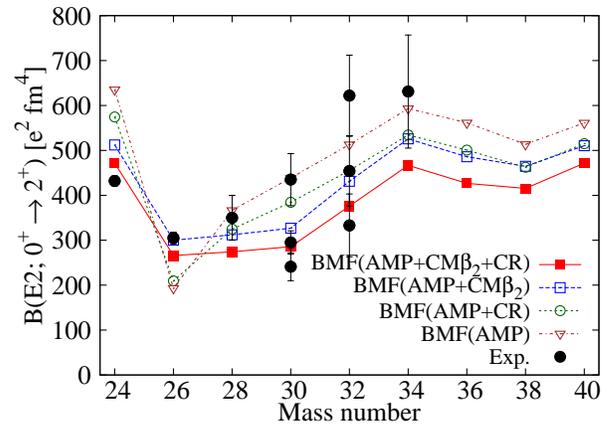}
\end{center}
\vspace{-12pt}
\caption
{(Color online)
$B(E2)$ transition probabilities of Mg isotopes.
Four results of BMF(AMP+CM$\beta_2$+CR), BMF(AMP+CM$\beta_2$),
BMF(AMP+CR), and BMF(AMP) calculations are compared
with experimental data (``Exp.'') from
Refs.~\cite{Moto95,Prit99,Chiste01,Raman01,Nied05,Iwasaki01}.
}
\label{fig:BE2}
\end{figure}

Figure~\ref{fig:BE2} shows the results for the $E2$ transition probabilities $B(E2;0^+ \rightarrow 2^+)$ together with experimental data where available.
No effective charge is required in our calculations.
Comparing the four calculations BMF(AMP+CM$\beta_2$+CR),
BMF(AMP+CM$\beta_2$), BMF(AMP+CR), and BMF(AMP),
it can be seen that both $\beta_2$ configuration mixing
and infinitesimal cranking yield non-negligible effects;
those of the former (latter) is about $8\%-28\%$ ($7\%-14\%$).
The combined effect on BMF(AMP) reduces the $B(E2)$ values
by about $15\%-35\%$ except for $^{26}$Mg,
which is predicted to be oblate in its ground state,
c.f. Fig.~\ref{fig:betahfbprjc}.
These effects are much larger on $B(E2)$ than on $r_{\rm m}$.
The BMF(AMP+CM$\beta_2$+CR) calculations
thus agree quite well with the measured $B(E2)$
when we consider that our calculations have no adjustable parameters.

%%%%%%%%%%%%%%%%%%%%%
%%%%%%%%%%%%%%%%%%%%

The $B(E2)$ transition probability within the rotational band
provides us with information on the quadrupole deformation parameter $\beta_2$.
Although it is not directly observable and a model is needed to determine
its value, this dimensionless quantity is quite useful and
often employed in the analysis of experimental data.
Reliable estimation of $\beta_2$ is thus meaningful
and is described in the Appendix; see Table~\ref{table:beta}.
In the Appendix, the absolute values of $\beta_2$
calculated from measured values of $B(E2)$ and $r_{\rm m}$ are
compared with those obtained by replacing $r_{\rm m}$ with
the empirical formula $r_{\rm m}^{\rm emp}=1.2A^{1/3}\sqrt{3/5}$~[fm],
c.f. Fig.~\ref{fig:beta2} in the Appendix.
The latter values overestimate the former by about 20\%
because $r_{\rm m}^{\rm emp}$ underestimates the experimental $r_{\rm m}$
by about 10\% as shown in Sec.~\ref{Matter radii}.
This clearly indicates that simultaneous analysis
of $B(E2)$ and $r_{\rm m}$ is important
for estimating the deformation precisely,
particular in order to avoid confusing the effects of deformation and expansion.

\subsection{Excitation energies}
\label{Excitation energies}

\begin{figure}[!htb]
\begin{center}
 \includegraphics[width=0.45\textwidth]{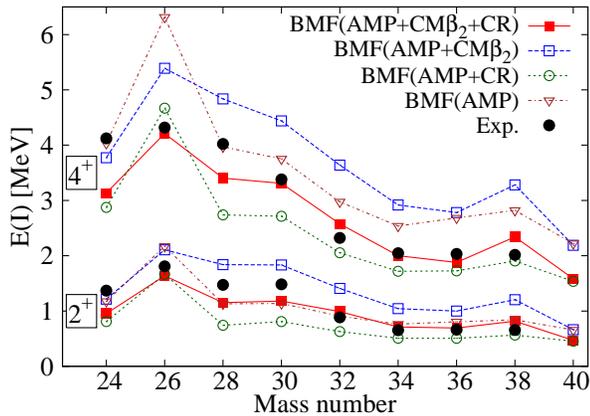}
\end{center}
\vspace{-12pt}
\caption
{(Color online)
Excitation energies $E(2^+)$ and $E(4^+)$ of Mg isotopes.
Four results of BMF(AMP+CM$\beta_2$+CR),
BMF(AMP+CM$\beta_2$), BMF(AMP+CR), and BMF(AMP) calculations are compared
with experimental data (``Exp.'')
from Refs.~\cite{NNDC, Door13} and references therein.
}
\label{fig:E24}
\end{figure}

The first $2^+$ and $4^+$ excitation energies $E(2^+)$ and $E(4^+)$
are shown in Fig.~\ref{fig:E24}.
The excitation energies from full BMF(AMP+CM$\beta_2$+CR) (filled squares)
agree quite well with the experimental data.
For $^{24}$Mg, however, the results  underestimate the experimental
values.  There are two possible reasons. One is that
the pairing gaps for both neutrons and protons vanish around
the projected minimum in the Gogny-D1S calculation,
which makes the moment of inertia larger.
Another possible reason is the effect of alpha clustering in $^{24}$Mg.
The results from BMF(AMP+CM$\beta_2$+CR) are rather different
from those in Ref.~\cite{RER02}
where the effect of cranking is not included and the excitation energies
are systematically larger, that is, the moments of inertia are smaller.
Our results are consistent with Ref.~\cite{RER15}
where the configuration mixing for both $(\beta_2,\gamma)$ deformation and
cranking frequency is taken into account. This agreement indicates
that the effect of triaxial deformation might not be very important,
at least for the low-lying states of the Mg isotopes.

%%%%%%%%%%%%%%%%%%%%%
%%%%%%%%%%%%%%%%%%%%

In contrast to the full BMF(AMP+CM$\beta_2$+CR) calculations,
the excitation energies from BMF(AMP+CR) (open circles) are
systematically lower than the experimental data.
The $\beta_2$-configuration mixing thus considerably increases
the excitation energies of the $2^+$ and $4^+$ states.
However, the results from BMF(AMP+CM$\beta_2$) (open squares)
greatly overestimate the experimental data.
This clearly shows that the effects of the time-odd components induced
by the infinitesimal cranking procedure are very important.
The effect of cranking induces $K$-mixing in the wave function
as shown in Eqs.~(\ref{eq:wf1})--(\ref{eq:wf4}), which, as a result, reduces
the energies of the $2^+$ and $4^+$ states but not the $0^+$ state in
which there is no $K$-mixing.   Therefore, the effects of
both $\beta_2$ configuration mixing and infinitesimal cranking
are important for reproducing excitation energies
for the $2^+$ and $4^+$ states of the ground-state rotational band.
The simplest BMF(AMP) almost reproduces the $2^+$ excitation energies
accidentally in these Mg isotopes.   The $4^+$ energies, however,
are overestimated considerably, and thus the BMF(AMP) calculation
does not describe the data very well.

%%%%%%%%%%%%%%%%%%%%%
\begin{figure}[!htb]
\begin{center}
 \includegraphics[width=0.45\textwidth]{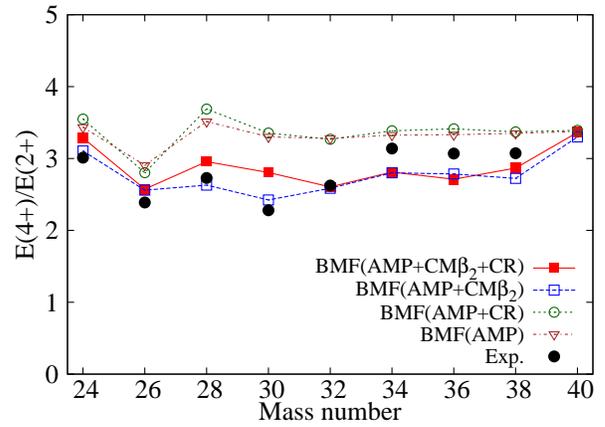}
\end{center}
\vspace{-12pt}
\caption
{(Color online)
Ratios $E(4^+)/E(2^+)$ for Mg isotopes.
The results of BMF(AMP+CM$\beta_2$+CR), BMF(AMP+CM$\beta_2$)
BMF(AMP+CR), and BMF(AMP) calculations are compared.
}
\label{fig:R24}
\end{figure}
%%%%%%%%%%%%%%%%%%%%

The ratios $E(4^+)/E(2^+)$ are shown in Fig.~\ref{fig:R24}.
The BMF(AMP+CM$\beta_2$) results without cranking
agree quite well with the experimental values,
although the energy is spaced too widely as can be seen in Fig.~\ref{fig:E24}.
Both the BMF(AMP) and BMF(AMP+CR) results
without $\beta_2$ configuration mixing
give values around the ideal rotational ratio of 3.3.
Inclusion of the effects of $\beta_2$-fluctuations reduces the ratio,
which clearly indicates that the effects of the $\beta_2$-fluctuation
included by the $\beta_2$ configuration mixing are very important
for describing the deviation from the ideal rotational behavior.
The ideal ratio of the vibrational motion is $E(4^+)/E(2^+)=2$,
which is closer to the experimental data for $^{26,30}$Mg.
However, comparing the results of BMF(AMP+CM$\beta_2$+CR)
and BMF(AMP+CM$\beta_2$), it can be seen that the infinitesimal cranking
does not affect the ratio very much.
The effect of infinitesimal cranking reduces the excitation energies,
but keeps the ratio constant.
That is, it only increases the moment of inertia
without changing the properties of the rotational motion.
The $\beta_2$-configuration mixing describes the deviation from
the ideal rotational motion and has a large influence
on the ratio $E(4^+)/E(2^+)$.

\begin{figure*}[!htb]
\begin{center}
%------------------------
\begin{tabular}{c}
 \begin{minipage}{0.36\hsize}
  \begin{center}
    \includegraphics[width=0.99\textwidth]{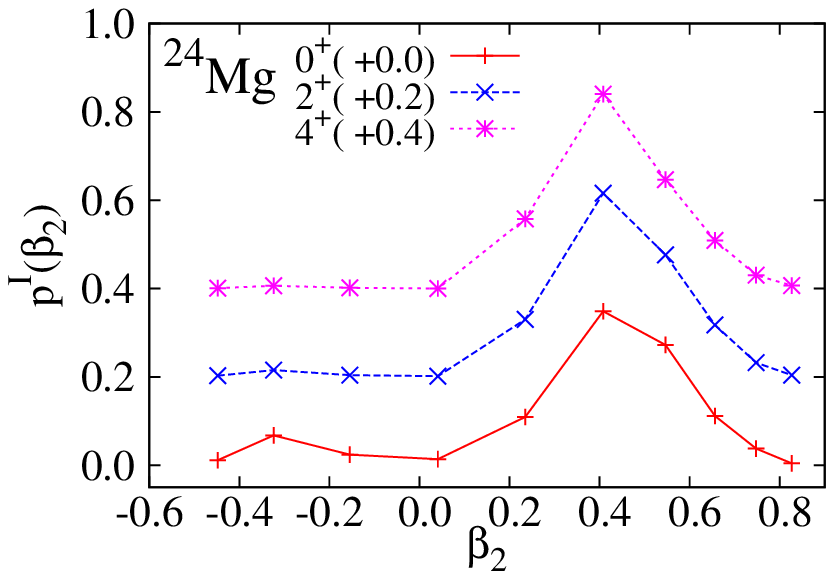}
  \end{center}
      \end{minipage}
 \begin{minipage}{0.36\hsize}
  \begin{center}
    \includegraphics[width=0.99\textwidth]{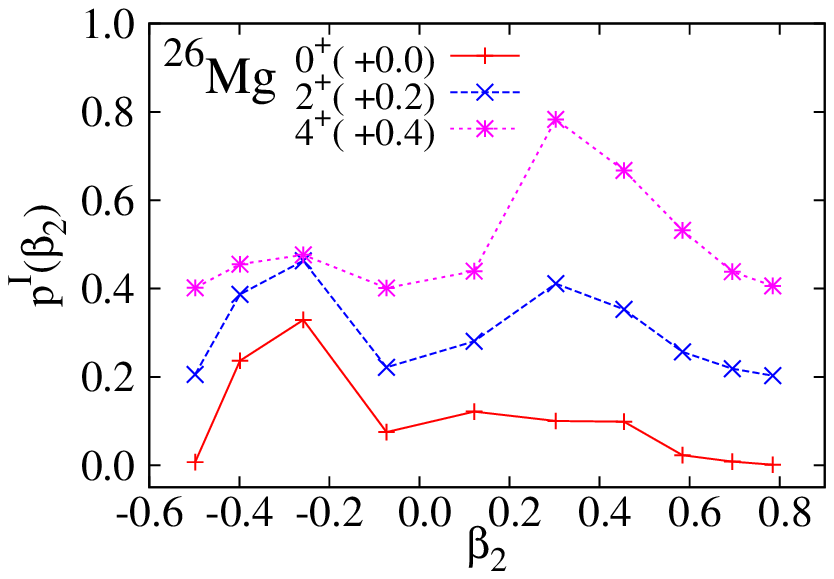}
  \end{center}
      \end{minipage}
\end{tabular}
%------------------------
\begin{tabular}{c}
 \begin{minipage}{0.36\hsize}
  \begin{center}
    \includegraphics[width=0.99\textwidth]{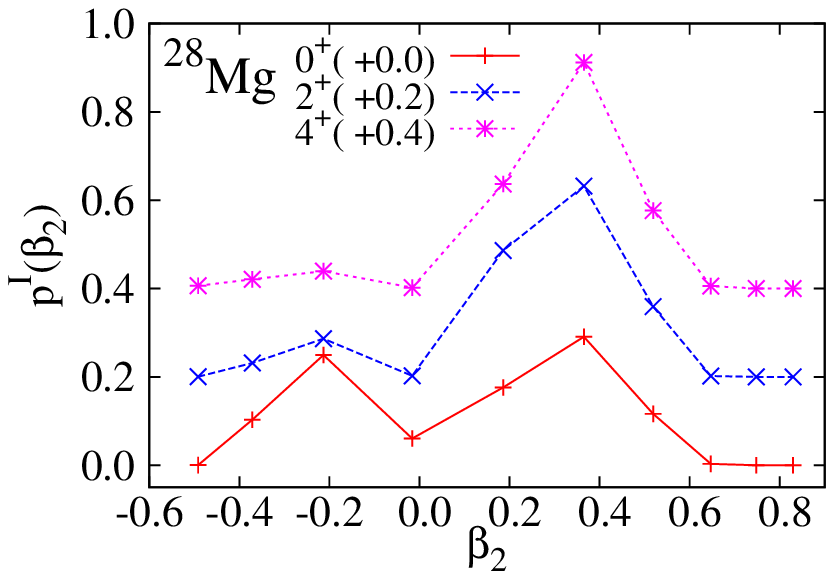}
  \end{center}
      \end{minipage}
 \begin{minipage}{0.36\hsize}
  \begin{center}
    \includegraphics[width=0.99\textwidth]{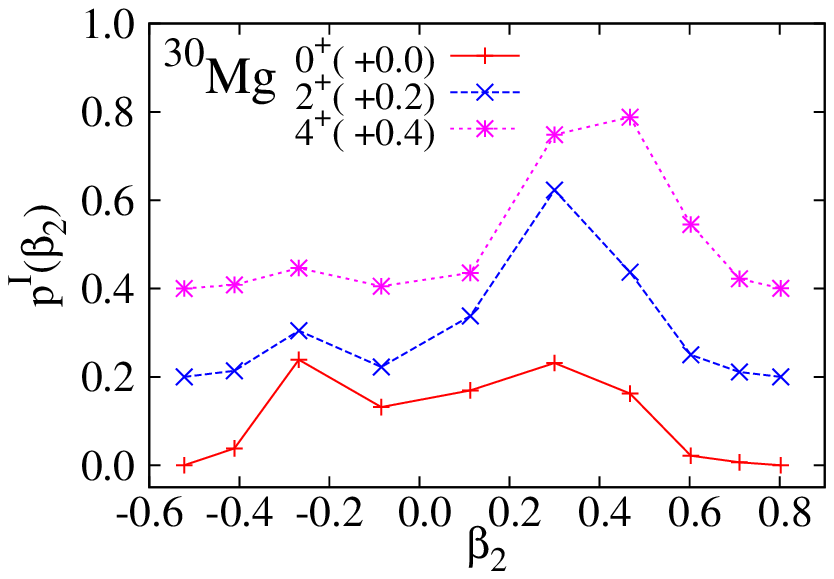}
  \end{center}
      \end{minipage}
\end{tabular}
%------------------------
\begin{tabular}{c}
 \begin{minipage}{0.36\hsize}
  \begin{center}
    \includegraphics[width=0.99\textwidth]{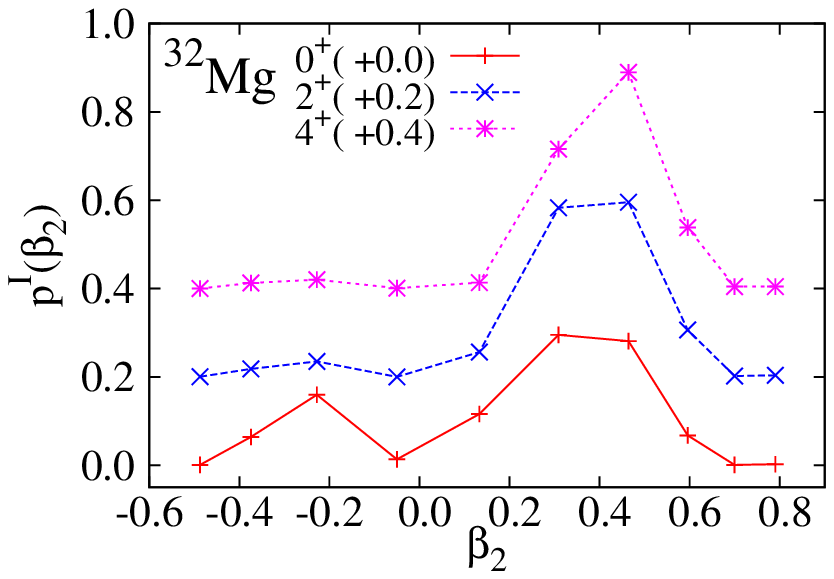}
  \end{center}
      \end{minipage}
 \begin{minipage}{0.36\hsize}
  \begin{center}
    \includegraphics[width=0.99\textwidth]{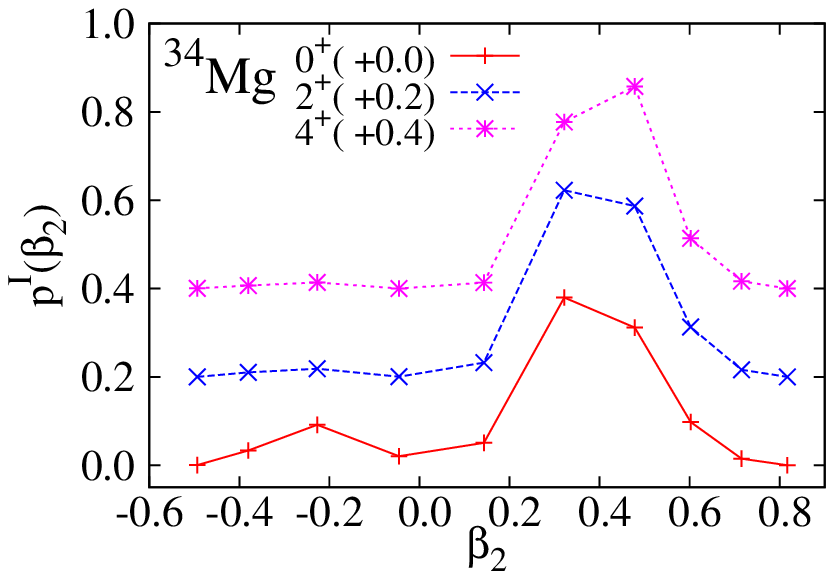}
  \end{center}
      \end{minipage}
\end{tabular}
%------------------------
\begin{tabular}{c}
 \begin{minipage}{0.36\hsize}
  \begin{center}
    \includegraphics[width=0.99\textwidth]{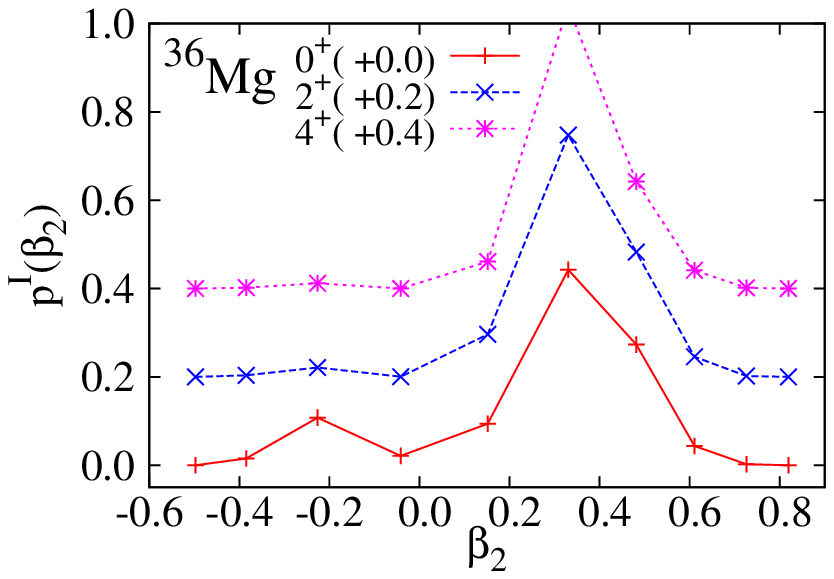}
  \end{center}
      \end{minipage}
 \begin{minipage}{0.36\hsize}
  \begin{center}
    \includegraphics[width=0.99\textwidth]{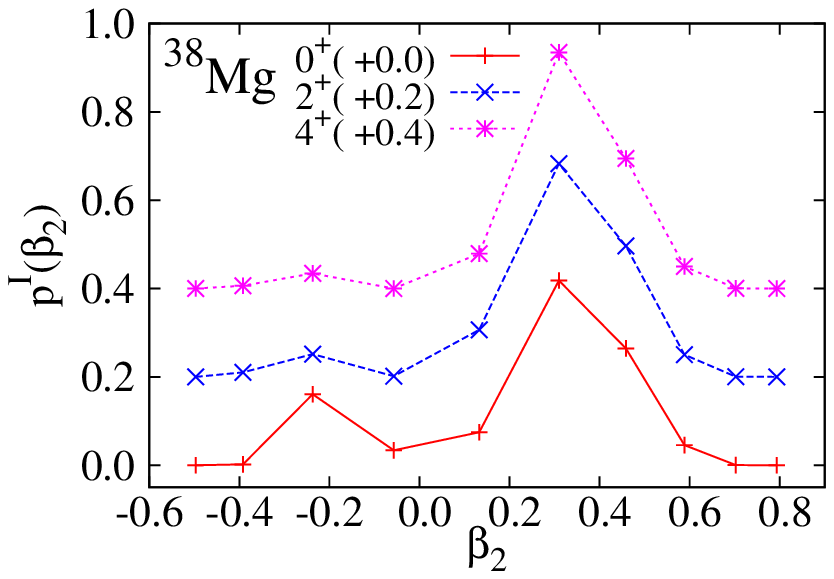}
  \end{center}
      \end{minipage}
\end{tabular}
%------------------------
\begin{tabular}{c}
 \begin{minipage}{0.36\hsize}
  \begin{center}
    \includegraphics[width=0.99\textwidth]{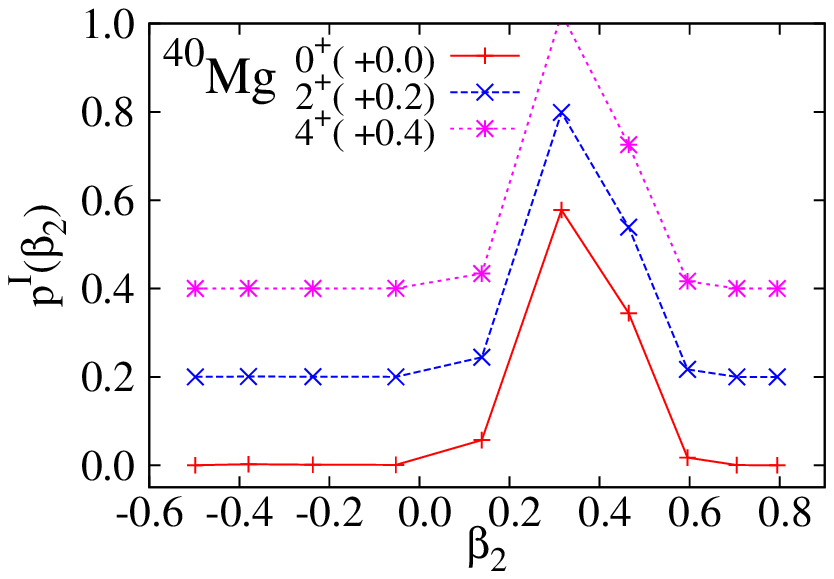}
  \end{center}
      \end{minipage}
 \begin{minipage}{0.36\hsize}
  \begin{center}
  \end{center}
      \end{minipage}
\end{tabular}
%------------------------
%\vspace{-3mm}
\caption
%$^{\arabic{Amass}}$
{(Color online)
Probability distributions from BMF(AMP+CM$\beta_2$+CR) calculations
for the $0^+, 2^+$ and $4^+$ states of Mg isotopes.
The corresponding distributions are shifted by $+0.2$ and $+0.4$
for the $2^+$ and $4^+$ states, respectively.
}
\label{fig:psidst_wt}
\end{center}
\end{figure*}

Figure~\ref{fig:psidst_wt} shows
the probability distributions of the $0^+, 2^+$ and $4^+$ states
by full BMF(AMP+CM$\beta_2$+CR) calculations.
For nuclei $^{24, 34, 40}$Mg, the distributions of
the $0^+$, $2^+$ and $4^+$ states are well located
in the prolate deformation side.
With the exception of $^{26}$Mg, the distributions of excited states are also located on the prolate side.
However, the distribution of the ground state is spread
across both the oblate and prolate sides for most nuclei.
In particular, an almost uniform distribution is seen in the wide range
around the spherical shape for the $0^+$ state in $^{30}$Mg.
This indicates that both the prolate and oblate configurations
contribute to the ground-state wave function.
For the nucleus $^{26}$Mg, an interesting transition in the distribution
occurs from the oblate to prolate side in which
the majority of mixing probabilities are
on the oblate side in the ground state whereas they are on the prolate side in the $4^+$ state.
Comparing the distributions of BMF(AMP+CM$\beta_2$+CR) calculations
with the those of BMF(AMP+CM$\beta_2$) calculations (not shown) shows that
they are quite similar to each other, although there is a slight change
in the $2^+$ and $4^+$ distributions for $^{26-30}$Mg nuclei.
Since infinitesimal cranking does not affect the $0^+$ state,
there is no effect on the $0^+$ distributions.

%-----------------------------
%\onecolumngrid

%%%%%%%%%%%%%%%%%%%%%
%    Fig.
%%%%%%%%%%%%%%%%%%%%%
%%%%%%%%%%%%%%%%%%%%

%\twocolumngrid
%-----------------------------

%Summary
\section{Summary}
\label{Summary}

In this paper, we simultaneously analyzed the three observables $r_{\rm m}$,
$B(E2)$, and $E(2^+)$ and $E(4^+)$ for $^{24-40}$Mg
using the BMF framework with
angular-momentum-projected configuration mixing
with infinitesimal cranking~\cite{TS16}.
The present BMF results are consistent with the results of
state-of-the-art BMF calculations in
Ref.~\cite{RER15} for $E(2^+)$ and $E(4^+)$ in $^{24-34}$Mg,
and with the experimental data.
Very recently, Ref.~\cite{TS16}
showed that infinitesimal cranking gives a practical and good description
of low-lying states.
The present BMF framework is thus considered to be practical and
rather reliable.
In fact, the present BMF calculations successfully reproduce all of the data for $r_{\rm m}$, $B(E2)$ and $E(2^+)$ and $E(4^+)$.
We thus conclude that the present BMF framework is quite useful
for data analysis, particularly for low-lying states.

We deduced the absolute value of the experimental deformation parameter
$\beta_2$ from measured values of $B(E2)$ and $r_{\rm m}$, and
present the resultant values in Table \ref{table:beta} in the Appendix.
Although the $\beta_2$ parameter is not directly measurable,
the present BMF framework is useful for extracting the values of $\beta_2$.
By comparing the values extracted from $B(E2)$ and $r_{\rm m}$
and from measured $B(E2)$ and the empirical formula
$r_{\rm m}^{\rm emp}=1.2A^{1/3}\sqrt{3/5}$~[fm], we show that
the latter overestimates the former by about 20\%,
since $r_{\rm m}^{\rm emp}$ underestimates $r_{\rm m}^{\rm exp}$ by about 10\%.
This clearly shows that simultaneous analysis of $r_{\rm m}$ and $B(E2)$
is quite important not only
for determining the deformation parameter precisely
but also for confirming the reliability of the theoretical framework.

We performed a detailed analysis by using the BMF framework which imposes the
restrictions that the system is axially symmetric and parity conserving.
The present BMF framework can take account of the following four effects,
separately:
(i) $\beta_2$ deformation,
(ii) change in $\beta_2$ by AMP
from the minimum of HFB energy surface to that of projected-energy surface,
(iii) configuration mixing for axially symmetric $\beta_2$ deformation,
and
(iv) inclusion of time-odd components by the cranking procedure.
Important effects are (i) and (ii) for $r_{\rm m}$ and (i) to (iv)
for both $B(E2)$ and $(E(2^+),E(4^+))$.
The effect (iv) especially reduces the values of
$E(2^+)$ and $E(4^+)$ without changing the ratio $E(4^+)/E(2^+)$.
The effect~(iv) thus enlarges the moment of inertia.
We thus propose that BMF(AMP) calculations
with effects (i) and (ii) are useful for analysis of the matter radii,
and full BMF(AMP+CM$\beta_2$+CR) calculations with effects (i) to (iv)
are useful for both the transition probabilities and
the excitation energies.

%Acknowledgement
%\section*{Acknowledgements}
\begin{acknowledgments}
This work is supported in part by
by Grant-in-Aid for Scientific Research
(Nos. 26400278 and 25$\cdot$4319)
from the Japan Society for the Promotion of Science (JSPS).
\end{acknowledgments}

%\noindent

%---------------------------------------------------------------
%\appendix
\appendix*
\section{Deformation parameter \bm{$\beta_2$}}

In this appendix, we discuss the values of the deformation parameter $\beta_2$
extracted from the measured $B(E2)$ and $r_m$.
For this purpose the quadrupole moment $Q_{20}$ in Eq.~(\ref{eq:beta2})
is expected to be related to $B(E2)$, for which the rotor model is necessary
(see, e.g., Ref.~\cite{RS80}).
However, the validity of the rotor model is questionable especially
for the case of smaller deformations in lighter systems.
Ref.~\cite{RB12} gives a condition for the validity of
the rotor model for $B(E2)$, Eq.~(\ref{eq:be2rot}); i.e.,
the angular-momentum fluctuation $\langle \Delta J^2 \rangle$
for the HFB state, from which the AMP is performed,
is larger than about 15$\hbar^2$.
In the present case, although the Mg isotopes are rather light,
the resultant deformations are relatively large, c.f. Table~\ref{table:beta}.
We checked that $\langle \Delta J^2 \rangle \ga 18-35\hbar^2$ for
the HFB states with deformation $|\beta_2|\approx 0.35-0.45$.
We therefore think that the rotor model can be applied quite safely.
Thus, according to the rotor model expression of $B(E2)$,
\begin{equation}
 B_{\rm rot}(\mathrm{E}2; I_i \rightarrow I_f )
 = \langle I_i 0 2 0 |I_f 0\rangle (Q^{(E)}_{20})^2,
 \label{eq:be2rot}
\end{equation}
where $Q^{(E)}_{20}$ is the electric quadrupole moment.
The mass quadrupole moment is extracted from the observed $B(E2)$ value
\begin{equation}
 |Q_{20}|=\frac{A}{Ze}\sqrt{B(E2;0^+ \rightarrow 2^+)},
\label{eq:Q20BE2}
\end{equation}
assuming the same deformation for neutrons and protons.
In this way, we extract the deformation parameters $|\beta_2|$
if experimental $r_{\rm m}$ are available,
as tabulated in Table~\ref{table:beta}.

%%%%%%%%%%%%%%%%%%%%%%%%%%%%%%
\renewcommand{\arraystretch}{1.25}
\begin{table}[!htb]
\begin{center}
\begin{tabular}{cccc}
\hline \hline
 Nuclide  &   $|\beta_2|$ &Error & $~~{\rm Ref.~for~} B(E2)$\\
\hline
 $^{24}$Mg  & 0.474  & 0.026 & \cite{Raman01}   \\
 $^{26}$Mg  & 0.409  & 0.014 & \cite{Raman01}   \\
 $^{28}$Mg  & 0.403  & 0.031 & \cite{Raman01}   \\
 $^{30}$Mg  & 0.336  & 0.023 & \cite{Nied05}    \\
 $^{30}$Mg  & 0.372  & 0.018 & \cite{Prit99}    \\
 $^{30}$Mg  & 0.452  & 0.031 & \cite{Chiste01}  \\
 $^{32}$Mg  & 0.410  & 0.036 & \cite{Moto95}    \\
 $^{32}$Mg  & 0.351  & 0.037 & \cite{Prit99}    \\
 $^{32}$Mg  & 0.480  & 0.035 & \cite{Chiste01}  \\
 $^{34}$Mg  & 0.445  & 0.046 & \cite{Iwasaki01} \\
\hline \hline
\end{tabular}
\caption
{
$|\beta_2|$ values obtained from measured values of $B(E2)$ and $r_{\rm m}$
by Eq.~(\ref{eq:beta2}) with~(\ref{eq:Q20BE2}).
}
\label{table:beta}
\end{center}
\end{table}
%%%%%%%%%%%%%%%%%%%%%%%%%%%%%%

%%%%%%%%%%%%%%%%%%%%%
\begin{figure}[!htb]
\begin{center}
 \includegraphics[width=0.45\textwidth]{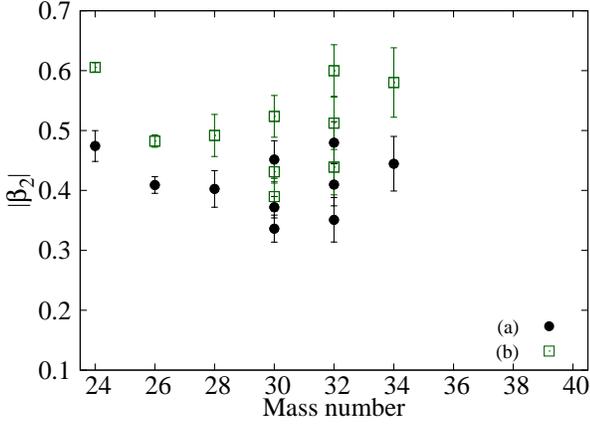}
\end{center}
\vspace{-12pt}
\caption
{(Color online)
Absolute values of (a) experimental and (b) empirical $\beta_2$ values.
Experimental $|\beta_2|$  (solid squares) are obtained
using Eqs.~(\ref{eq:beta2}) and~(\ref{eq:Q20BE2})
from measured values of $B(E2)$ and $r_{\rm m}$,
while empirical $|\beta_2|$  (open squares) are evaluated
from measured $B(E2)$ and empirical radii
$r_{\rm m}^{\rm emp}=1.2A^{1/3}\sqrt{3/5}$~[fm].
}
\label{fig:beta2}
\end{figure}
%%%%%%%%%%%%%%%%%%%%

%%%%%%%%%%%%%%%%%%%%%
\begin{figure}[!htb]
\begin{center}
 \includegraphics[width=0.45\textwidth]{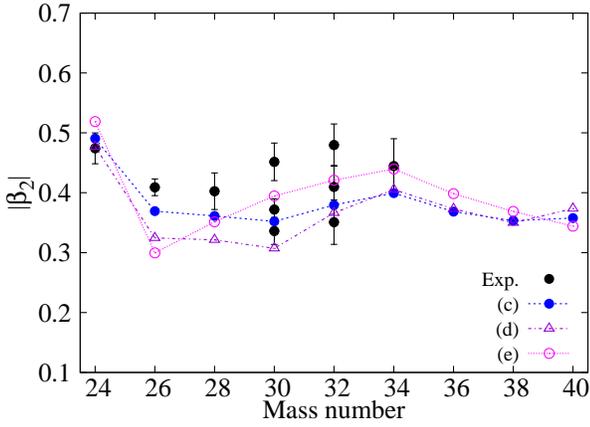}
\end{center}
\vspace{-12pt}
\caption
{(Color online)
Absolute values of three kinds of theoretical $\beta_2$ parameters:
(c) $|\beta_2|$ obtained
using Eqs.~(\ref{eq:beta2}) and~(\ref{eq:Q20BE2})
from the results of full BMF(AMP+CM$\beta_2$+CR)
 for $B(E2)$ and $r_{\rm m}$ (solid circles);
(d) mean $\beta_2$ values (${\bar \beta}_2$) of
Eq.~(\ref{eq:beta2m}) (open triangles); and (e) $\beta_2$ values
that give a minimum of projected energy surface (open circles).
These values are compared with the experimental $|\beta_2|$ values.
}
\label{fig:beta2-2}
\end{figure}
%%%%%%%%%%%%%%%%%%%%

Figure~\ref{fig:beta2} compares (a) the absolute values of $\beta_2$
extracted from measured $B(E2)$ and $r_{\rm m}$(solid circles)
with (b) values from measured $B(E2)$ and empirical radius $1.2 A^{1/3}$
(open squares), respectively, for $A=24-34$ Mg isotopes.
The $|\beta_2|$ values assuming the empirical radii (b) overestimate
those extracted from the measured radii (a) by about 20\%
because the empirical radii underestimate the measured values
by about 10\% as shown in Fig.~\ref{rms_exp}.
This clearly indicates that simultaneous analysis of
$B(E2)$ and $r_{\rm m}$ is important for estimating the deformation reliably.
Figure~\ref{fig:beta2-2} shows the $A$ dependence of
three kinds of theoretical $\beta_2$ parameters:
(c) $|\beta_2|$ deduced from the results of full BMF(AMP+CM$\beta_2$+CR)
calculations for $B(E2)$ and $r_{\rm m}$ (solid circles);
(d) mean $\beta_2$ values (${\bar \beta}_2$)
defined by Eq.~(\ref{eq:beta2m}) (open triangles);
and (e) $\beta_2$ values corresponding to the minimum of
the $0^+$ energy surface obtained by BMF(AMP) calculations (open circles),
c.f., Sec.~\ref{Energy surfaces} for details.
The $|\beta_2|$ values of BMF(AMP+CM$\beta_2$+CR) (c)
reproduce the experimental data quite well.
The $|\beta_2|$ (d) also account for the experimental data
well except for $^{26-30}$Mg.
For $^{26-30}$Mg, the probability distributions are spread quite widely
across both the oblate and prolate sides of $\beta_2$ with non-negligible
probability in small deformation as shown in Fig.~\ref{fig:psidst_wt}.
This leads to effective reduction of deformation, and consequently
the $|\beta_2|$ (d) slightly underestimates the experimental data.
The $|\beta_2|$ (e) corresponding to the minimum of
the projected energy surface
is slightly different from BMF(AMP+CM$\beta_2$+CR),
which shows the effect of configuration mixing.

%%--------------------------------------------------------------------%%
%%                           References                               %%
%%--------------------------------------------------------------------%%

\end{document}